\documentstyle[astrobib,psfig]{mn-ab}

\title{On the cosmological distance and redshift
between any two objects}

\author[J. Liske et al.]{J.~Liske\thanks{E-mail: jol@phys.unsw.edu.au}\\ 
	School of Physics, University of New South Wales, Sydney 2052,
	Australia}

\date{Accepted
...... Received .....}

\pagerange{\pageref{firstpage}--\pageref{lastpage}}

\pubyear{2000}

\newcommand{\be}{\begin{equation}}
\newcommand{\ee}{\end{equation}}
\newcommand{\bea}{\begin{eqnarray}}
\newcommand{\eea}{\end{eqnarray}}
\newcommand{\eref}[1]{(\ref{#1})}
\renewcommand{\d}{{\rm d}}

\psfigurepath{plots/}

\begin{document}

\label{firstpage}
\maketitle

\begin{abstract}
We discuss the problem of how to calculate the distance between two
cosmological objects given their redshifts and angular separation on
the sky. Although of a fundamental nature, this problem and its
solution seem to lack a detailed description in the literature. We
present a new variant of its solution and quantitatively assess the
most commonly used approximation.
\end{abstract}


\section{Introduction}
In cosmology and extragalactic astronomy one frequently needs to
calculate the distance between two objects given their redshifts and
their angular separation on the sky. As larger and larger cosmological
volumes are probed by wide field redshift surveys such as 2dF and the
Sloan Digital Sky Survey, or by QSO absorption line studies, the
effects of non-Euclidean geometry become increasingly important. In
the past, authors have frequently relied on approximations when
calculating the distance between two cosmological objects, presumably
for calculative ease. Such approximations are valid only for small
distances and are particularly useful when examining or highlighting
the feasibility of geometrical means to measure cosmological
parameters as was done e.g.\ by \citeN{Alcock79}, \citeN{Phillipps94}
and \citeN{Popowski98}. However, many practical applications are not
limited by computing time and since an exact solution to the problem
exists the approximations seem unnecessary. These applications include
the construction of the real-space two-point correlation function of
various objects such as galaxies (e.g.\ \citeNP{Yoshii93}) and QSO
absorption systems (\citeNP{Williger96}; \citeNP{Dinshaw96}), as well
as studies of the effect of local sources of ionizing radiation on
their surrounding intergalactic medium (e.g \citeNP{Fernandez95}). The
latter application actually requires knowledge of not only the
distance between two objects but also of the redshift experienced by a
photon travelling from one object to another.

Given the fundamental nature of this problem we feel a detailed,
explicit treatment is called for. In this short article we present a
new variant of the solution to the distance problem
(Sections~\ref{mysol} and \ref{results}), discuss its relation to
existing approaches (Section~\ref{oldsol}) and finally investigate the
validity of the most frequently used approximation
(Section~\ref{comp}).

For clarity and brevity we have limited ourselves in this paper to
homogeneous Friedmann (zero-pressure) cosmologies with no cosmological
constant ($\Lambda = 0$). The inclusion of $\Lambda$ renders some of
the explicit expressions non-analytical and thus (in the context of
this paper) unnecessarily complicates matters.

\section{The distance between any two objects} \label{mysol}
We begin by writing the familiar Robertson-Walker line element as:
\be \label{rwm}
\d s^2 = -c^2 \d t^2 + a^2(t)\left[\d \chi^2 + \Sigma ^2(\chi)(\d \theta^2 + 
\sin^2\theta \: \d \phi^2)\right], 
\ee
where
\be
\Sigma(\chi) = \left\lbrace
\begin{array}{lccl}
\sin\chi \qquad & k & = & +1 \\
\chi & k & = & 0 \\
\sinh\chi & k & = & -1.
\end{array} \right.
\ee
Putting the Earth at the origin of the coordinate system one can use
this metric and the Friedmann equations to calculate the distance from
Earth (at $\chi = 0$) to an object at redshift $z$, corresponding to a
comoving coordinate $\chi$,
\bea \label{r}
r & = & a_0 \Sigma[\chi(z)] \nonumber \\
& = & \frac{c}{H_0 q_0^2} \: \frac{1}{1 + z} \nonumber \\
& & \times \left[q_0 z + (q_0-1)\left(\sqrt{1 + 2 q_0 z} - 1\right) \right]
\eea
(e.g.\ \citeNP{Misner}), where $a_0$, $H_0$ and $q_0$ are the scale
factor, Hubble and deceleration parameters at the present epoch
(subscript $0$).

\begin{figure}
\psfig{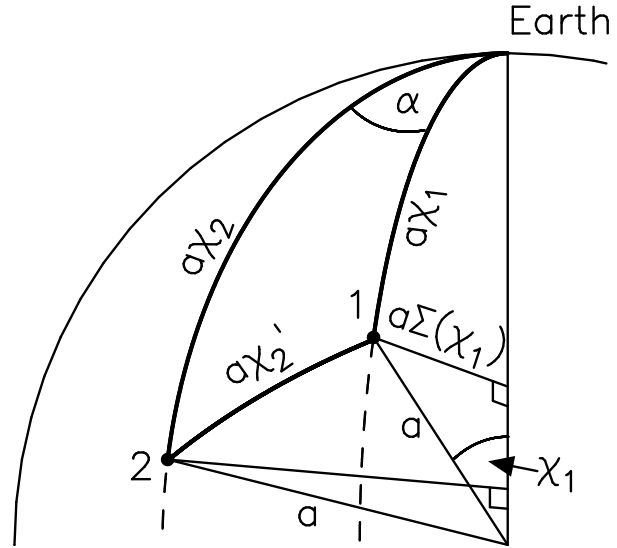}
\caption{In the case $k = +1$ objects 1 and 2 and Earth form a
triangle with geodesic sides (thick lines) on the surface of a 2d
sphere of radius $a$. The Earth is at the origin of the coordinate
system (`north pole'). From the centre of the sphere the angle between
Earth and object 1 is $\chi_1$. A photon emitted by object 1 towards
Earth travels along the geodesic connecting the two (thick line). The
length of this path (= distance between object 1 and Earth) is $a
\chi_1$. $a \Sigma(\chi_1) = a \sin \chi_1$ is the distance from
object 1 to the central axis of the sphere. Essentially this is the
angular diameter distance or luminosity distance (modulo factors of
$1+z$) from Earth to object 1. $\chi_2'$ is the unknown angle between
objects 1 and 2 at the centre of the sphere and $a \chi_2'$ is the
unknown length of the connecting geodesic.}
\label{coord}
\end{figure}

Now consider an object 1 (the `receiver') observed on Earth today at
$z_1$ and an object 2 (the `emitter') at $z_2$ separated by an angle
$\alpha$ on the sky (cf.\ Fig.~\ref{coord}). Object 2 emits a photon
towards object 1 which is received by object 1 at the same time as
object 1 emits the photon we receive from it today (i.e.\ at the epoch
corresponding to $z_1$). What is the distance, $r_2'$, between these
two objects at the time of the photon reception and what is the
redshift, $z_2'$, of the photon as observed by object 1?

An observer at object 1 would write equation \eref{r} as
\bea \label{r2'}
r_2' & = & a_1 \Sigma(\chi_2') \nonumber \\
& = & \frac{c}{H_1 q_1^2} \: \frac{1}{1 + z_2'} \nonumber \\
& & \times \left[q_1 z_2' + (q_1-1)\left(\sqrt{1 + 2 q_1 z_2'} - 1\right) 
\right],
\eea
where $\chi_2'$ is the comoving coordinate distance between objects 1 and 2
(cf.\ Fig.~\ref{coord}) and
\bea
a_1 & = & \frac{a_0}{1 + z_1}, \nonumber \\ 
H_1 & = & H_0 (1 + z_1) \sqrt{1 + 2 q_0 z_1},\\
q_1 & = & q_0 \frac{1 + z_1}{1 + 2 q_0 z_1} \nonumber
\eea
are the scale factor, Hubble and deceleration parameters at the time
object 1 emitted the photons we receive today. Thus we see that the
problems of calculating $r_2'$ and $z_2'$ are equivalent since
knowledge of one provides knowledge of the other via equation
\eref{r2'}. Here we choose to find $z_2'$.

Note that $r_2'$ does {\em not} describe in general the shortest
distance between objects 1 and 2 along a $t = {\rm const}$
hypersurface of spacetime.  Nevertheless, in many applications $r_2'$
is the quantity of interest. E.g.\ when considering the radiative
effect of a QSO on a nearby object one needs the luminosity distance
between the two which is given by $r_{{\rm L}2}' = r_2' (1 +
z_2')$. In any case, it is most practical to solve the problem for
$z_2'$ and then calculate the distance required for a given
application from $z_2'$.

As we are dealing with three points (Earth and objects 1 and 2) in a
3-dimensional space (described by the spatial part of the metric,
equation \ref{rwm}) it seems intuitive that it must be possible to
reduce the problem to two dimensions as one can always find a 2d
hypersurface that contains all three points. Since we are interested
in measuring 3d distances the hypersurface should be chosen such that
the distance between any two points of the surface (as measured along
the surface) is identical to the 3d distance between the same two
points. We call such surfaces totally geodesic. It is clear that a
totally geodesic hypersurface containing a given set of three points
can only be constructed from the geodesics connecting the three
points. Since these geodesics are unique there can be only one such
surface.

In equation \eref{rwm} we introduced a polar coordinate system $(\chi,
\theta, \phi)$. Clearly, the hypersurface described by $\phi = {\rm
const}$ is totally geodesic. (Note that the surfaces $\chi = {\rm
const}$ and $\theta = {\rm const}$ are not.) Since the curvature of
the 3d space under consideration is constant, one can generate {\em
all} totally geodesic hypersurfaces from any given one by mere
translations and rotations. Therefore, for a given set of three points
there must exist a coordinate system $(\tilde \chi, \tilde \theta,
\tilde \phi)$ such that $\tilde \phi = {\rm const}$ describes the
unique totally geodesic hypersurface containing these three points.
Since this new coordinate system can be constructed from the old one by
translation and rotation, the form of the metric in this new system is
identical to equation \eref{rwm}. Restraining this metric to the
$\tilde \phi = {\rm const}$ hypersurface we have
\be
\d l^2 = a^2 \left[\d \tilde \chi^2 + \Sigma^2(\tilde \chi) 
\d \tilde \theta^2 \right].
\ee
Thus we can see that the triangle Earth--object 1--object 2 lies
either on a 2d sphere, a plane or a 2d hyperboloid ($k = +1, 0, -1$)
embedded in 3d Euclidean space. This triangle has geodesic sides
$\chi_1, \chi_2,$ and $\chi_2'$ (connecting the two objects) and the
angle $\alpha$ at Earth. The case $k = +1$ is shown in
Fig.~\ref{coord}.

The objective is now to express the unknown side $\chi_2'$ in terms of
the known sides $\chi_1$, $\chi_2$ and the angle $\alpha$. We first
note that all formulae of Euclidean trigonometry have corresponding
formulae in spherical and hyperbolical trigonometry. These can be
expressed simultaneously for all three curvatures using $\Sigma$. In
particular, we can generalize the half-angle formulae in this way and
use them to show that
\bea \label{master}
\Sigma^2\left(\frac{\chi_2'}{2}\right) & = & 
\Sigma^2\left(\frac{\chi_2 + \chi_1}{2}\right) \sin^2\frac{\alpha}{2}
\nonumber \\ 
& & \mbox{} + \Sigma^2\left(\frac{\chi_2 - \chi_1}{2}\right) 
\cos^2\frac{\alpha}{2}.
\eea
This is a more compact and symmetrical version of the generalized
cosine rule (see Section \ref{oldsol}).

Using the same methods that were employed in the derivation of equation
\eref{r} we can relate the right-hand side of equation \eref{master}
to $z_1$ and $z_2$:
\be \label{2pm1}
a_0 \Sigma\left(\frac{\chi_2 \pm \chi_1}{2}\right) = 
\frac{c}{H_0 q_0} \: \frac{1}{\sqrt{(1 + z_1) (1 + z_2)}} \: 
\frac{P_\pm}{2},
\ee
where
\bea
P_+ & = & \frac{1}{q_0} \left[ (q_0 - 1) 
\left(\sqrt{1 + 2 q_0 z_1} 
+ \sqrt{1 + 2 q_0 z_2} - 1\right) \right. \nonumber \\ 
& & \left. \mbox{}+ \sqrt{(1 + 2 q_0 z_1) (1 + 2 q_0 z_2)} - q_0 \right]
\eea
and
\be
P_- = \left( \sqrt{1 + 2 q_0 z_2} - \sqrt{1 + 2 q_0 z_1} \right).
\ee
Furthermore, setting $\chi_1 = z_1 = 0$ it follows from equation 
\eref{2pm1} that
\be
a_0 \Sigma\left(\frac{\chi_2}{2}\right) = 
\frac{c}{H_0 q_0} \: \frac{1}{\sqrt{1 + z_2}} 
\: \frac{1}{2} \left( \sqrt{1 + 2 q_0 z_2} - 1 \right)
\ee
and for an observer at object 1
\be
a_1 \Sigma\left(\frac{\chi_2'}{2}\right) = 
\frac{c}{H_1 q_1} \: \frac{1}{\sqrt{1 + z_2'}} \:
\frac{1}{2} \left( \sqrt{1 + 2 q_1 z_2'} - 1 \right),
\ee
thus relating the left-hand side of equation \eref{master} to
$z_2'$. Solving
\be
P = \frac{1}{2} \frac{1}{\sqrt{1 + z_2'}} \left( \sqrt{1 + 2 q_1 z_2'} 
- 1 \right)
\ee
for $z_2'$ yields (positive solution)
\be \label{z_2'}
z_2' = \frac{2 P^2}{(q_1 - 2 P^2)^2} \left( 1 + q_1 - 2 P^2 +
\sqrt{\frac{q_1^2}{P^2} + 1 - 2 q_1} \right),
\ee
where
\bea \label{P2}
P^2 & = & \left(\frac{a_1 H_1 q_1}{c}\right)^2 \Sigma^2
\left(\frac{\chi_2'}{2}\right)
\nonumber \\
& = & \frac{1}{4} \; \frac{1 + z_1}{1 + z_2} \; \frac{1}{1 + 2 q_0 z_1}
\left(P_+^2 \sin^2\frac{\alpha}{2} + P_-^2 \cos^2\frac{\alpha}{2} \right).
\eea
Note that $z_2'$ does not depend on $H_0$.

\subsection{$\Lambda \not= 0$}
For completeness we now briefly consider the case $\Lambda \not= 0$.
Equation \eref{master} is of course still valid.  However, in relating
this equation to the observables $z_1$, $z_2$ and $z_2'$ we have made
use of the fundamental relationship between an object's comoving
radial coordinate and its redshift,
\be
\chi(z) = \frac{c}{a_0 H_0} \int_{1}^{1+z} 
\left[ \Omega_0 x^3 + (1 - \Omega_0) x^2 \right]^{-\frac{1}{2}} \d x,
\ee
which in turn is derived from the Friedmann equations. For $\Lambda
\not= 0$ these take on a different structure and instead of the above
relationship we have
\be
\chi(z) = \frac{c}{a_0 H_0} \int_{1}^{1+z} 
\left[ \Omega_0 x^3 + (1 - \Omega_0 - \lambda_0) x^2 + \lambda_0
\right]^{-\frac{1}{2}} \d x,
\ee
where $\lambda_0 = \Lambda c^2 / 3 H_0^2$. Unfortunately this integral
is non-analytical but \citeN{Kayser97} developed an efficient method
(which also accommodates inhomogeneity) to compute $\Sigma(\chi)$
numerically.

Note that other density contributions with unusual equations of state
can be dealt with in the same way. All that needs to be done is to
establish $\chi(z)$ or, equivalently, $\Sigma[\chi(z)]$. In principle,
one could then proceed to use equation \eref{master} to find $z_2'$
and $r_2'$. However, when $\chi(z)$ is not available analytically it
is probably more practical to use the generalized cosine rule (see
Section \ref{oldsol}, equation \ref{cosine}) instead of equation
\eref{master}. In any case, one is still left with the problem of
inverting $\Sigma[\chi(z)]$ in order to find $z_2'$. However, since we
can compute
\be
\frac{\d \Sigma}{\d z} = \sqrt{1 - k \Sigma^2(z)} \: \frac{\d \chi}{\d z}
\ee
it should be possible to employ an efficient root finding algorithm
for this task.

\section{Results} \label{results}
\begin{figure}
\psfig{file=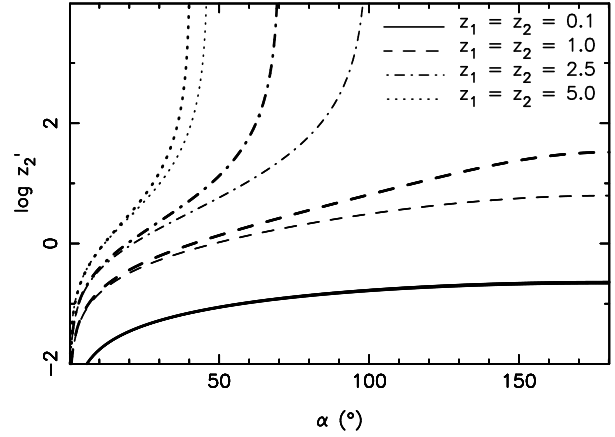,width=0.45\textwidth,angle=-90,silent=}
\caption{$z_2'$ as a function of $\alpha$ (angle on the sky between
receiver and emitter), $z_1 = z_2$. Thick lines are for $q_0 = 0.5$,
thin lines for $q_0 = 0.15$.}
\label{z_alpha}
\end{figure}

\begin{figure}
\psfig{file=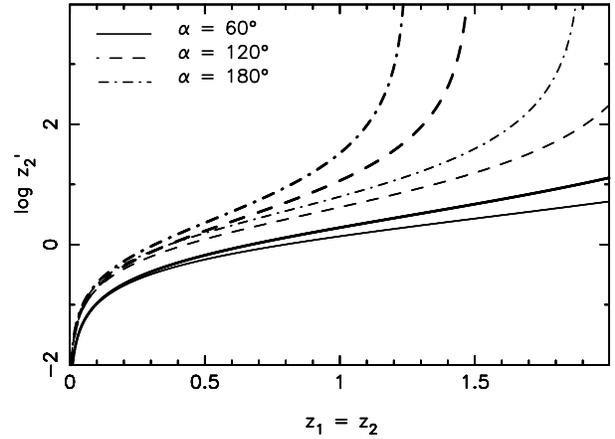,width=0.45\textwidth,angle=-90,silent=}
\caption{$z_2'$ as a function of $z_1 = z_2$. Thick lines are for
$q_0 = 0.5$, thin lines for $q_0 = 0.15$.}
\label{z_z1z2}
\end{figure}

\begin{figure}
\psfig{file=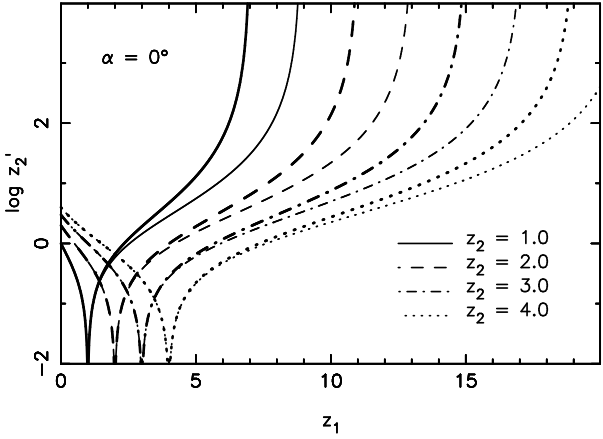,width=0.45\textwidth,angle=-90,silent=}
\caption{$z_2'$ as a function of $z_1$ (redshift of receiver), $\alpha
= 0$. Thick lines are for $q_0 = 0.5$, thin lines for $q_0 =
0.15$. Since $\alpha = 0$, $z_2'$ is independent of cosmology for $z_1
< z_2$.}
\label{z_z1}
\end{figure}

\begin{figure}
\psfig{file=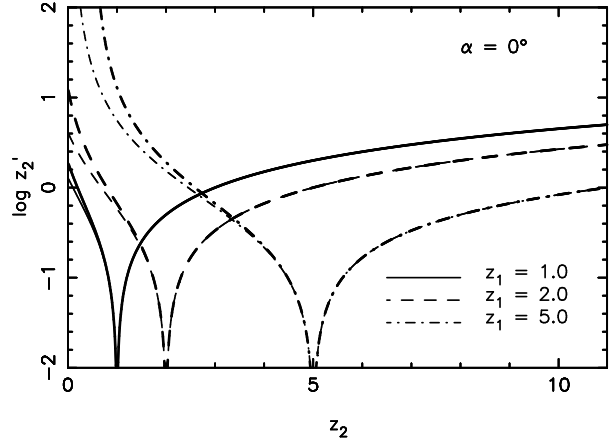,width=0.45\textwidth,angle=-90,silent=}
\caption{$z_2'$ as a function of $z_2$ (redshift of emitter), $\alpha
= 0$. Thick lines are for $q_0 = 0.5$, thin lines for $q_0 =
0.15$. Since $\alpha = 0$, $z_2'$ is independent of cosmology for $z_1
< z_2$.}
\label{z_z2}
\end{figure}

\begin{figure}
\psfig{file=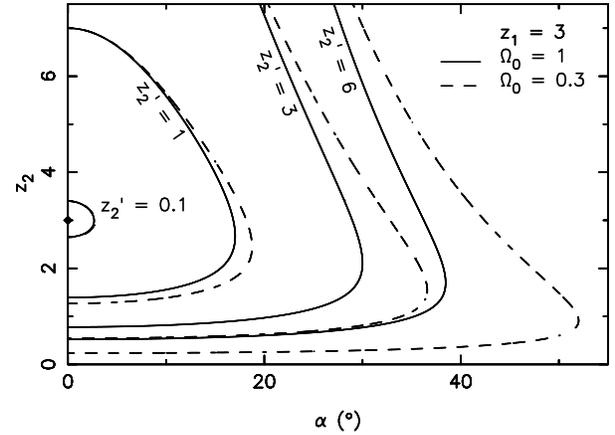,width=0.45\textwidth,angle=-90,silent=}
\caption{$z_2'$ as a function of $z_2$ and $\alpha$ (emitter
position). The receiver is at $z_1 = 3$ and $\alpha = 0$ (marked
by a diamond). The contours are lines of constant $z_2'$. Solid
lines are for $q_0 = 0.5$, dashed lines for $q_0 = 0.15$.}
\label{z_z2alpha}
\end{figure}

As cosmologists are often used to thinking in terms of redshift rather
than distance, we show the result of the above calculations in
Figs.~\ref{z_alpha}--\ref{z_z2alpha} in terms of $z_2'$, the
redshift of object 2 as seen by object 1.

Figs.~\ref{z_alpha} and \ref{z_z1z2} explore the special case $z_1 =
z_2$. In Fig.~\ref{z_alpha} we show $z_2'(\alpha)$ with $z_1 = z_2$
fixed at various values. The thick lines show the case of a flat
universe ($\Omega_0 = 2 q_0 = 1$) and the thin lines show the case of
an open universe with $\Omega_0 = 0.3$. Note that for large values of
$z_1 = z_2$ there is some $\alpha_\infty$ such that $z_2' \rightarrow
\infty$ for $\alpha \rightarrow \alpha_\infty$. This is the particle
horizon of object 1 at the epoch corresponding to $z_1$. At that time,
light emitted from objects separated from object 1 by angles $>
\alpha_\infty$ has not had time to reach object 1 since the Big Bang
(ignoring inflation). In Fig.~\ref{z_z1z2} we fix $\alpha$ at various
angles and show $z_2'$ as a function of $z_1 = z_2$. The case $\alpha
= 180\degr$ is often incorrectly used in undergraduate physics
textbooks (e.g.\ \citeNP{Halliday93}, p.\ 1128--1129) as an example of
how to add velocities in Special Relativity, a method which will give
a wrong result for $z_2'$.

In Figs.~\ref{z_z1} and \ref{z_z2} we consider the special case
$\alpha = 0$ and plot $z_2'$ as a function of $z_1$ and $z_2$
respectively. Thick lines again represent a flat universe, thin lines
an open universe. Whenever $z_1 < z_2$ (and $\alpha = 0$), $z_2'$
is of course given by
\be
1 + z_2' = \frac{1 + z_2}{1 + z_1}
\ee
which is the only case where $z_2'$ is independent of the cosmological
model, since in this case the time of emission of the photon received
by object 1 is the same as the time corresponding to $z_2$.  Note that
although $\chi_2'(z_1, z_2, \alpha) = \chi_2'(z_2, z_1, \alpha)$ (cf.\
equation \ref{cosine}), the same does {\em not} hold for $z_2'$ (cf.\
equations \ref{z_2'} and \ref{P2}).  Fig.~\ref{z_z2} also provides
the solution to an interesting thought experiment. What is the
redshift, $z_{\rm refl}$, of a photon emitted by ourselves ($z_2 = 0)$
which was reflected back to us by a comoving mirror at $z_1$? The
answer is
\be
1 + z_{\rm refl} = (1 + z_1) [1 + z_2'(z_1, z_2 = 0)].
\ee

Finally we plot in Fig.~\ref{z_z2alpha} lines of constant $z_2'$ as a
function of $z_2$ and $\alpha$ for a receiver at $z_1 = 3$. In this
plot the flat and open cosmologies are represented by the solid and
dashed lines respectively.

\section{Relation to other solutions} \label{oldsol}
We are aware of two original solutions to the problem in the literature,
\citeN{Peacock99}, p.\ 71, and \citeN{Osmer81}. Neither of them consider
$z_2'$.

Instead of equation \eref{master}, Peacock considers the generalized 
cosine rule which directly gives $\Sigma(\chi_2')$. Introducing the
cosine equivalent of $\Sigma(\chi)$,
\be
\Xi(\chi) = \sqrt{1 - k \Sigma^2(\chi)},
\ee
we have
\be
\Xi(\chi_2') = \Xi(\chi_1) \: \Xi(\chi_2) + k \Sigma(\chi_1) \Sigma(\chi_2)
\cos \alpha
\ee
which can be written as
\bea \label{cosine}
\Sigma^2(\chi_2') & = & \Sigma^2(\chi_1) \: \Xi^2(\chi_2) + \Sigma^2(\chi_2)
\: \Xi^2(\chi_1) \nonumber \\
& & \mbox{} + k \Sigma^2(\chi_1) \Sigma^2(\chi_2) \sin^2 \alpha \nonumber \\
& & \mbox{} - 2 \Sigma(\chi_1) \Sigma(\chi_2) \: \Xi(\chi_1) \: \Xi(\chi_2) 
\cos \alpha.
\eea
When only $r_2'$ is needed the use of this equation seems more
practical than our solution presented in Section \ref{mysol}. However,
when using the generalized cosine rule the analogues of equations
\eref{z_2'} and \eref{P2} are more complicated so that in cases where
$z_2'$ is (also) of interest, e.g.\ when calculating the luminosity
distance $r_{{\rm L}2}' = r_2' (1 + z_2')$, the new solution is to be
preferred.

Osmer's solution may be considered the most rigorous as it is based on
a general result of differential geometry in maximally symmetric
spaces. \citeN{Weinberg72}, p.\ 413, showed how to transform to a
coordinate system which has been `quasitranslated'. Osmer uses this
equation to transform from a coordinate system in which objects 1 and
2 have particularly simple coordinates (see discussion in Section
\ref{mysol}) to one where the origin has been translated from Earth to
object 1. The result is then given by
\bea
\Sigma^2(\chi_2') & = & \Sigma^2(\chi_2) \sin^2 \alpha \nonumber \\
& & \mbox{} + [\Sigma(\chi_2) \: \Xi(\chi_1) \cos \alpha
- \Sigma(\chi_1) \: \Xi(\chi_2)]^2.
\eea
This equation of course reduces to equation \eref{cosine} and thus 
the comments made there apply equally to Osmer's solution.

\section{Validity of approximation} \label{comp}
\begin{figure}
\psfig{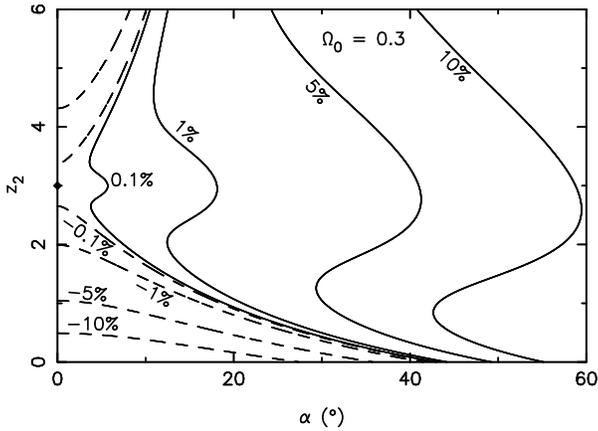}
\caption{The contours are lines of constant $\frac{a_0 \Delta
\chi_2'}{a_0 \chi_2'}$, the fractional error made when using equation
\eref{chiapprox} for $z_1 = 3$ (marked by a diamond) and $q_0 =
0.15$. The contour levels are indicated.}
\label{deltachi}
\end{figure}

Probably the most common approximation for the comoving
distance is (e.g.\ \citeNP{Yoshii93}; \citeNP{Phillipps94})
\be \label{chiapprox}
a_0^2 \, \chi_2'^2 \approx a_0^2 \, \chi_{{\rm A}2}'^2 =
a_0^2 \, \Sigma^2[\chi(\overline{z})] \: \alpha^2 + 
\left[a_0 \frac{\d \chi}{\d z}(\overline{z})\right]^2 \Delta z^2,
\ee
where $\Delta z = z_1 - z_2$ and $\overline{z} = (z_1 + z_2)/2$ and
\be
a_0 \frac{\d \chi}{\d z}(z) = \frac{c}{H(z)}.
\ee
In Fig.~\ref{deltachi} we plot the fractional error made when
using equation \eref{chiapprox},
\be
\frac{a_0 \Delta \chi_2'}{a_0 \chi_2'} = \frac{a_0 \chi_{{\rm A}2}'
- a_0 \chi_2'}{a_0 \chi_2'},
\ee
against $z_2$ and $\alpha$ for object 1 at $z_1 = 3$ and for $q_0 =
0.15$. We see that the approximation gives both too large and too
small distances (solid and dashed contours) depending on the position
in the $z_2$-$\alpha$ plane. Note the `ridges' along which the
approximation incidentally gives the correct distance.

The special case $\Delta z = 0$ deserves some further attention as it
corresponds to the well-known angular diameter problem: what is the
length, L, of a rod at $z$ which subtends an angle $\alpha$ as seen
from Earth? Commonly, the answer is given as
\be \label{ang_dist}
L_{\rm A} = \alpha \: r_{\rm A} = \frac{\alpha \: a_0 \Sigma(\chi)}{1 + z}
\ee
where $r_{\rm A}$ is the angular diameter distance. However, strictly
speaking $L_{\rm A}$ is the length of the line $\chi = {\rm const}$
connecting the two ends of the rod at the epoch $z$, which is {\em
not} the shortest distance between the ends. The length of the
geodesic connecting the two ends is given by $a\chi_2'$ which we can
derive from equation \eref{master}. Setting $\chi_1 = \chi_2 = \chi$
and $z_1 = z_2 = z$ we arrive at
\bea \label{L}
L & = & a \chi_2' \nonumber \\
& = & \frac{a_0}{1 + z} \: 2 \: \Sigma^{-1}\left[\Sigma(\chi) \:
\sin\frac{\alpha}{2} \right].
\eea
The fractional error made when using equation \eref{ang_dist},
$\frac{\Delta L}{L} = \frac{L_{\rm A} - L}{L},$
is of course the same as in \eref{chiapprox} for $\Delta z = 0$ since
the extra factor $(1 + z)$ cancels out. However, comparing equations
\eref{ang_dist} and \eref{L} it is particularly easy to see that
approximations such as \eref{chiapprox} actually contain two
approximations: (a) small angle and (b) neglect of curvature (for $k =
\pm 1$). E.g.\ $\frac{\Delta L}{L}$ is independent of $z$ in the flat
case but is larger and varies with $z$ for the open case.

\section*{Acknowledgments}
We thank C.~Lineweaver for useful discussions.

\label{lastpage}

\end{document}